\documentclass[aps,pre,preprint,superscriptaddress,amsmath,amssymb,nofootinbib,showpacs]{revtex4}

\usepackage{amsmath}
\usepackage{graphicx}
\usepackage{amssymb}
\usepackage{upgreek}

\makeatletter

\begin{document}

\title{Chemotaxis of ciliated microorganisms: with and without noise}

\author{Ruma Maity}\thanks{rumamaity@phy.iitkgp.ac.in}
\affiliation{Department of Physics, Indian Institute of Technology Kharagpur, Kharagpur, India}

\author{P.S. Burada}\thanks{Corresponding author: psburada@phy.iitkgp.ac.in}
\affiliation{Department of Physics, Indian Institute of Technology Kharagpur, Kharagpur, India}
\affiliation{Center for theoretical studies, Indian Institute of Technology Kharagpur, Kharagpur, India}

\date{\today}

\begin{abstract}

Biological systems like ciliated microorganisms are capable to respond to the external chemical gradients, a process known as chemotaxis 
which has been studied here using the chiral squirmer model. 
This theoretical model considers the microorganism as a spherical body with an active surface slip velocity. 
In presence of a chemical gradient, the internal signaling network of the microorganism is triggered due to binding 
of the ligand with the receptors on the surface of the body. 
Consequently, the coefficients of the slip velocity get modified resulting in a change in the path followed by the body. 
We observe that the strength of the gradient is not the only parameter which controls the dynamics of the body 
but also the adaptation time play a very significant role in the success of chemotaxis of the body. 
Path of the body is smooth if we ignore the discreteness in the ligand-receptor binding which is stochastic in nature. 
In presence of the later, the path is not only irregular but the dynamics of the body changes. 
We calculate the mean first passage time, by varying strength of the chemical gradient and adaptation time, to 
investigate the success rate of chemotaxis.\end{abstract}

\pacs{
47.15.G-, 47.63.Gd, 87.17.Jj, 78.20.Bh    
}

\maketitle

\section{Introduction}

Chemotaxis is the movement of a single cell or multicellular organism in response to a chemical stimulus.
It is ubiquitous in several biological processes, e.g., 
fertilization where the chemoattractants released by the egg guide the sperm cell to reach it \cite{Friedrich}, 
early development of multicellular organisms  \cite{Hadwiger},
wound healing \cite{tanya}, embryogenesis \cite{martin},
food finding for the survival of the species \cite{wang} etc. 
In recent past, artificial chemotaxis is an emerging field of interest \cite{pulak,lagzi} where 
the synthetic systems are designed such that they can sense the chemical gradients and execute the programmed action.
The later is very useful in many technological and medical applications like artificial fertilization, cancer treatment \cite{sahari}, etc.
Also, the artificial bodies are being prepared to sense gradients of light \cite{dai}, temperature \cite{bickel} etc.  

In nature, at a microscopic level, sperm cells and microorganisms like \textit{E. Coli, Dictyostelium, Paramecium, 
Tetrahymena thermophila, Amoeba proteus} etc. exhibit chemotaxis \cite{Larsen,nebl,Jennings,Shamloo,Nakatani,Houten,almagor,korohoda}.
Chemotaxis of \textit{E.Coli} and sperm cells is well studied \cite{Friedrich,dev,samanta, Larsen, Julicher,hussain,lu,yoshida, jikeli,pichlo,darszon}.
The sperm cells have flagella which generate wavelike motion to propel the body in the forward direction. The flagella contains receptors which can bind with the chemoattractants molecules leading to the activation of an internal signaling network. 
This sequentially changes the intracellular $Ca^{2+}$ concentration of the body which in return changes 
the beating pattern and swimming frequency of flagella \cite{Friedrich, Julicher,Alvarez}. 
Thus, the body can move towards or away from the chemical source \cite{miravete}. On the other hand, 
it has been observed experimentally that the \textit{E.Coli} uses run and tumble strategy to move.
Also the rotation of \textit{E. Coli}'s flagella depends on the 
type of chemotactic agent; for attractants flagella rotates in counter clockwise direction 
while for repellants it rotates in clockwise direction \cite{Larsen}. 

However, chemotaxis of other systems, in particular, the ciliated microorganisms has not been explored much.
Interestingly, how a ciliated microorganism like \textit{Paramecium} senses the gradient is not very clear.
While some experimental evidences suggest that the presence of specific binding sites on the ciliary membrane of 
\textit{Paramecium} \cite{mike} is accountable for its response to a specific chemical stimulus, 
others pointed out that the receptors are on the cell membrane of the organism \cite{oami}. 
Not only \textit{Paramecium} but majority of ciliates have receptors either as a primitive feature or as a consequence of evolution \cite{antipa}. 
Also, recently Shah \emph{et. al.} \cite{Shah} and others \cite{Preston} reported that the motile cilia are also able to 
perform sensory functions which changes the earlier paradigm that only primary cilia has receptors on it.

In reality, the chemotactic signalling process is not free from noise \cite{Friedrich} as 
the binding of chemoattractants to the receptors is a discrete random process.
It results in a fluctuation in receiving and sensing the stimulus.
This random process can be referred to as a chemical noise, which may affect the behavior of the microorganism. 
To minimize the effect of noise the body needs to adjust its internal parameters minutely. 
Though noise seems to be a disadvantageous situation which can delay the body's arrival at the target, sometimes its presence proves to be helpful.
For example, if the body is stuck just in the middle of two equally strong chemoattractive sources, the noise will help the body to break the symmetry 
and move towards either of them. 
On the other hand, since size of the most microorganisms is of the order of $\mu m$, 
the body is too large for the thermal noise to be effective \cite{Elegeti, Battle}.

Most of the ciliated microorganisms propel in the fluid due to synchronous beating of cilia leading to metachronal waves at their surface.
This induces an active surface slip. 
The motion of ciliated microorganisms have been studied earlier using the well known squirmer model \cite{Lighthill,Blake,Reviews, Lauga, Ishikawa}, 
a sphere with an axisymmetric surface slip. This simple squirmer exhibit translational motion only. 
However, in general ciliated microorganisms exhibits not only translational motion but also body rotation \cite{Crenshaw} 
which gives rise to helical motion, e.g. \textit{Strombidium sulcatum} \cite{tom}, \textit{Paramecium} in confined geometry \cite{jana} etc. 
Recently, Burada \emph{et. al.} have introduced a more general squirmer model called the chiral squirmer, which takes into account 
the body rotations \cite{Burada,Pak}. The rotation rate in addition to the translational velocity results in a helical path of the squirmer. 
In this paper, we consider the chiral squirmer model to study the chemotaxis of ciliated microorganisms both in absence and in presence of noise.

The paper is organized as follows. In Sec. II, we describe our model system. 
In Sec. III, we study the chemotaxis of our model system by applying both the linear and the radial chemical gradients. 
The influence of noise in the process of chemotaxis is investigated in Sec. IV. 
We present our main conclusions in Sec. V.

\section{The chiral squirmer model}

In general, ciliated microorganisms are low Reynolds number swimmers \cite{Lighthill} 
and obey the Stoke's equation \cite{Happel} given by,
\begin{equation}
 \eta\nabla^2 \mathbf{v} = \nabla p\,,
\end{equation}
where $\eta$ is the viscosity of the fluid around the body, $\mathbf{v}$ is the velocity field generated by the body in the surrounding fluid, 
and $p$ is the corresponding pressure field. 

In the chiral squirmer model, the effect of metachronal waves generated by cilia of 
the mircoorganism is taken care by the active slip velocity \cite{Burada}. 
In addition, if we consider that the body is non-deformable then the radial component of slip velocity is zero and we are left 
with tangential components only to describe the active slip on the surface. 
Hence, the effective slip velocity on the surface of the body is defined as \cite{Burada}, 
\begin{align}
\mathbf{v}_s = \beta \nabla_s Y_1^0  + \displaystyle{\sum \limits_{m=-1}^1} \left(\gamma_{m}\mathbf{e_r}\times\nabla_sY_1^m \right)\,,
\label{eq:slip}
\end{align}
where $\nabla_s$ is the surface gradient operator given by, 
$\nabla_s = \mathbf{e_ \theta} \frac{\partial }{\partial \theta} + (1/\sin\theta) \mathbf{e_ \phi} \frac{\partial}{\partial \phi}$ 
and $Y_1^m$ are the spherical harmonics.
The unit vectors $\mathbf{e_r}$, $\mathbf{e_ \theta}$ and $\mathbf{e_ \phi}$ are along $r$, $\theta$ and $\phi$ directions, respectively. 
The parameters $\gamma_m$ are the complex slip coefficients defined as $\gamma_m = \gamma_m^r + i\,m \gamma_m^i$. 
Note that in the slip velocity, the higher modes, e.g., $Y^m_l (l>1)$ have not been included as they do not contribute to the propulsion of the body. 
Thus, they are not relevant for the current study.

The velocity, rotation rate, and dissipative power of the chiral squirmer can be obtained directly from the slip velocity \cite{Stone}. They are given by- 
\begin{align}
\mathbf{U} & = \frac{2}{3} \beta\, \textbf{t}\,, \label{eq:vel}\\
\mathbf{\Omega} & = \frac{\gamma_{1}^r}{a}\,\textbf{n} + \frac{\gamma_{1}^i}{a}\,\textbf{b} + \frac{\gamma_0^r}{a}\,\textbf{t}\,, \label{eq:omega}\\
P & = 12 \pi \eta a(|\mathbf{U}|^2 + \frac{4a^2 }{9}|\mathbf{\Omega|}^2) \label{eq:power}\,,
\end{align}
where (\textbf{n}, \textbf{b}, \textbf{t}) is the body frame of reference and $a$ is the radius of the chiral squirmer. 
Note that in the following we set $\gamma_1^i = 0$ for simplicity. 
Equations of motion of the body can be obtained by solving the force- and torque balance equations,
\begin{align}
\dot{\textbf{r}} = \textbf{U}, \,\,\,\, 
\dot{\textbf{n}} = \mathbf{\Omega}\times \textbf{n},\,\,\,\,
\dot{\textbf{b}} = \mathbf{\Omega}\times \textbf{b},\,\,\,\,
\dot{\textbf{t}} = \mathbf{\Omega}\times \textbf{t}\,.
\end{align}
These coupled equations can be solved analytically to obtain position $\mathbf{r}(t)$ of the chiral squirmer \cite{Suwan, Lowen},
\begin{align}
\label{eq:path}
\mathbf{r(t)}  = \mathbf{r_0} + 
\frac{(\mathbf{\Omega}\times \mathbf{U}) \times \mathbf{\Omega}}{\left|\mathbf{\Omega}\right|^3} \sin\,(\left|\mathbf{\Omega}\right|t) + 
\frac{(\mathbf{\Omega}\cdot \mathbf{U})}{\left|\mathbf{\Omega}\right|^2} \mathbf{\Omega} \,t 
+ \frac{(\mathbf{\Omega}\times \mathbf{U})}{\left|\mathbf{\Omega}\right|^2} (1- \cos\,(\left|\mathbf{\Omega}\right|t)) \,.
\end{align}
Depending on the angle between $\mathbf{U}$ and $\mathbf{\Omega}$ path of chiral squirmer is 
either a straight line; for $\mathbf{U} \parallel \mathbf{\Omega}$, 
or a circle; for $\mathbf{U} \perp \mathbf{\Omega}$, or a helix; for other angles \cite{Burada,Pak}. 

\section{Chemotaxis in absence of noise}

In presence of a chemical gradient, the chemoattractants bind to the receptors on the microorganism. 
This binding is called activation and triggers the internal signaling network. 
The body then senses the ligand or rather the relative change in the ligand concentration ($\Delta c/ c$) in the vicinity \cite{ned}. 
Here, $c$ is the local stimulus level. 
This sensitivity is a function of ligand-binding affinity \cite{victor} which decreases with increasing concentration of ligand \cite{ralph}. 
As a result, the body gets adapted to the external stimulus \cite{springer}.
The output of the ligand-receptor binding event is the entry of the $Ca^{2+}$ into the cells along the cilia, 
resulting in a change in the velocity and rotation rate of the microorganism \cite{Friedrich,Martin}. 
This forces the body to change its natural path and follow the gradient. 
The density of the intracellular $Ca^{2+}$ depends on the ambient ligand concentration. 
After sometime the chemoattractant is unlaced from the receptor and the internal $Ca^{2+}$ is 
removed from the cells which are then depolarised again. 
This process is called deactivation. 
Following the deactivation, the second chemoattractant appears to the receptor. 
This give rise to a new $Ca^{2+}$ signal.
Note that, signals due to different chemoattractants are independent and they do not superimpose on each other \cite{Kashikar}.
Thus the system first adapts and then relaxes in presence of the chemical gradient by a series of activation and deactivation processes.
This helps the squirmer to move either towards or away from the target. 
The dynamics of adaptation and relaxation can be captured by the following coupled equations \cite{Barkai},
\begin{subequations}
\begin{equation}
\sigma \dot{a_b} = p_b (s_b + s) - a_b\,,
\end{equation}
\begin{equation}
\mu \dot{p_b} = p_b (1-a_b)\,,
\end{equation}
\label{eq:adt_relax}
\end{subequations}
where $\sigma$ is the relaxation time, $\mu$ is the adaptation time, $a_b(t)$ is the 
dimensionless output variable which is related to the internal $Ca^{2+}$ concentration,  
$p_b (t)$ is the dynamic sensitivity related to adaptation and $s_b(t)$ arises from 
the background activity of receptors in absence of the chemical stimulus. 
The chemoattractant has a dimension of concentration. 
Note that, these equations are valid for weak concentration gradients only.
Under a constant stimulus $s(t) = S_c$, the system reaches a steady state for which $a_b = 1$ and $p_b = \frac{1}{s_b + S_c}$. 
Therefore, dynamic sensitivity ($p_b$) maintains an inverse relation with $s_b$ and $s$. 
With increasing stimulus level $p_b$ decreases.
Since the steady state value of the output variable $a_b(t)$ is independent of the stimulus $S_c$, the system is totally adaptive.

In general, in ciliated microorganisms, the presence of external stimuli changes the beating pattern of cilia.
That is manifested in the current model by modifying the slip coefficients as follows,
\begin{subequations}
\begin{equation}
  \beta = \beta^{(0)} + \beta^{(1)}[a_b(t) - 1] \,,
\end{equation}
\begin{equation} 
  \gamma_0^{r} = \gamma_0^{r(0)} + \gamma_0^{r(1)}[a_b(t) - 1] \,,
\end{equation}  
\begin{equation}  
  \gamma_1^{r} = \gamma_1^{r(0)} + \gamma_1^{r(1)}[a_b(t) - 1] \, , 
 \label{betas}
\end{equation} 
 \end{subequations}
where $\beta^{(0)}$, $\gamma_0^{r(0)}$, and $\gamma_1^{r(0)}$ are the unperturbed slip coefficients.
The parameters $\beta^{(1)}$, $\gamma_{0}^{r(1)}$, and $\gamma_{1}^{r(1)}$ 
are due to external chemotactic stimulus. 
For the sake of simplicity, in the following we use dimensionless units.
In particular, we scale lengths by radius of the squirmer $a$,
time by $t_0 = a/U_0$, pressure by $ p_0 =(\mu U_0)/a$ ($U_0$ is the unperturbed velocity of the squirmer),
and adaptation by $p_c$ (the steady state value).
In the following, we study the chemotaxis of a chiral squirmer in presence of both the linear and the radial chemical gradients.

\subsection{Linear chemical gradient}

The linear chemical concentration is defined by \cite{Friedrich},
\begin{equation}
c(\mathbf{r}) = c_0 + \mathbf{c^\prime_1}\cdot \mathbf{r}\,,
\label{eq:l_gradient}
\end{equation}
where the constant $c_0$ is uniform chemoattractant concentration,
$\mathbf{c^\prime_1}$ is the chemical gradient, i.e., $\mathbf{c^\prime_1} = \mathbf{i}\,c_1 = \nabla c(\mathbf{r})$,
with the strength $c_1$ and $\mathbf{r} = \mathbf{i} x + \mathbf{j} y + \mathbf{k} z$ is the position vector. 
Thus, the chemotactic stimulus reads $\mathbf{s}(t) = \mathbf{c}(\mathbf{r}(t))$ \cite{Friedrich}.

\begin{figure}[t]
\centering
\includegraphics[scale = 0.4]{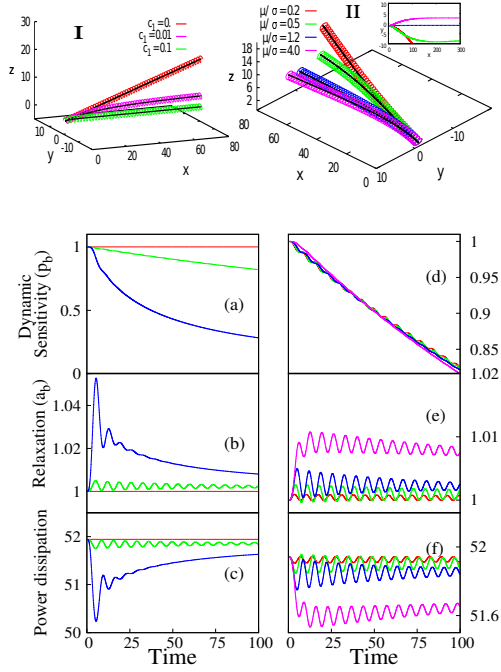}
\caption{(Color online) 
(I) Depicts path of the squirmer for various strengths of chemical gradient ($c_1$) for the linear case.
For higher values of $c_1$ squirmer makes a quick turn towards the direction of the chemical gradient. 
(a)-(c) Shows the corresponding behaviors of $p_b$, $a_b$, and power dissipation as a function of time, respectively.
The other parameters are $\mu/\sigma = 1.2$, $s_b = 0$, $\beta^{(0)} = 3/2$, $\beta^{(1)} = \beta^{(0)}/10$, $\gamma_{0}^{r(0)} = 0.2$, 
$\gamma_{0}^{r(1)} = 2$, $\gamma_{1}^{r(0)} = 0.9$, $\gamma_{1}^{r(1)} = -2$.
For higher value of $c_1$, $p_b$ decays faster.
With increasing $c_1$, the rate of binding of chemoattractants onto the body is higher which is reflected in $a_b$ 
where we can see that higher peaks corresponds to higher gradients. 
Note that for a constant stimulus $a_b$ maintains its steady state value which is one.
(II) Shows path of the squirmer for various adaptation times for the linear case, and 
(d)-(f) are the corresponding behaviors of $p_b$, $a_b$, and power dissipation as a function of time, respectively.
For this, we have chosen $c_1 = 0.01$.}
\label{fig:lin_nonoise}
\end{figure}

In absence of a chemical gradient, the squirmer moves in a helical path with a velocity and rotation rate given by 
Eq.~\ref{eq:vel} and \ref{eq:omega}, respectively \cite{Burada}. 
In presence of a chemical gradient, the squirmer changes its natural path and moves towards 
the chemical gradient, see Fig.~\ref{fig:lin_nonoise}(I).
Thus, its sensitivity $(p_b)$ starts to decrease (see fig.\ref{fig:lin_nonoise}(a)) because it 
maintains an inverse relation with the local stimulus level. 
Steeper the gradient more rapid is the fall of $p_b$ with time indicating that the body is advancing towards higher chemical concentration region quickly. 
The peaks in $a_b$ are analogous to the loading of $Ca^{2+}$ into the cells. 
The peak height depends on the strength of gradient, see fig.\ref{fig:lin_nonoise}(b). 
Since higher gradient increases binding rate i.e., 
the chemoattractant density at the receptors, $Ca^{2+}$ increases accordingly. 
When the sensitivity becomes very low, $a_b$ tends to reach to its unperturbed state, see fig.\ref{fig:lin_nonoise}(b). 
While the peaks in fig.\ref{fig:lin_nonoise}(b) are associated with the turning of the body towards the gradient, 
the decreasing part of $a_b$ is associated with the alignment of the body to the direction of the gradient, see fig.\ref{fig:lin_nonoise}(I). 
For simplicity, we have assumed that the linear velocity of the body is only slightly perturbed by the presence of the gradient. 
On the contrary, the direction of the $\Omega$ decides the trajectory of the body. 
Therefore, the components of $\Omega$ get modified greatly under the gradient. 

Note that, the adaptation is a robust property while the adaptation time $\mu$ \cite{Barkai} which is analogous to the memory of the microorganism is not. 
An optimum value of $\mu$ exists for which the chemotaxis is most favorable \cite{Nikita, Victor}.
The key of successful chemotaxis lies in comparing the past and the present stimulus levels and respond instantaneously. 
Body with shorter $\mu$ forgets the past stimulus quickly whereas with longer $\mu$ remembers for a longer time. 
For shorter $\mu$, the internal signaling network is reset by removing the memory associated with the past signal 
before the body could compare it to the present stimulus level. 
As a result, it is difficult for the body to follow the gradient as shown in fig. \ref{fig:lin_nonoise}(II).
On the other hand, higher $\mu$ delays the body's turning towards the appropriate direction because the network is not 
able to reset itself by erasing the response due to the past signal rapidly.
This leads to higher $Ca^{2+}$ peaks which takes longer time to decay as compared to that with shorter $\mu$, 
see figs. \ref{fig:lin_nonoise}(e). 
Also, for higher $\mu$ before the body could completely drain out the $Ca^{2+}$ from the cells, it experiences the new stimulus level by following the gradient. 
Therefore, we get a series of peaks in higher $a_b$ values, see fig.\ref{fig:lin_nonoise}(e). 
Hence, only for an optimum value of $\mu$ the body align in the direction of gradient ($x$ direction) quickly, see fig. \ref{fig:lin_nonoise}(II).
Injection of $Ca^{2+}$ into the cells increases the linear velocity but slows down the rotation rate of the body.
The dynamic sensitivity $p_b$ decays over time for different $\mu/\sigma$ values qualitatively but with different rates. 
This implies, that the power dissipation is less for higher $\mu$, see fig. \ref{fig:lin_nonoise}(f).

\subsection{Radial chemical gradient}

The radial chemical concentration is defined as,
\begin{equation}
c(\mathbf{r}) = \frac{c_r}{r}  \,,
\label{rad_c}
\end{equation} 
where $c_r$ is a constant which depends on the diffusivity of chemoattractants, i.e. 
the rate at which chemoattractants are released from the chemical source 
and $r = \sqrt{x^2 + y^2+z^2}$ is the distance between the squirmer and the chemical source which is placed at the origin. 
$c(r)$ can be expressed in dimensionless form as, $\tilde{c}(\mathbf{r}) =  c(\mathbf{r}) / c_{0r}(\mathbf{r}) $, 
where $c_{0r}(\mathbf{r}) = c_r/r_0$ and $r_0 = \sqrt{x_0^2 + y_0^2 + z_0^2}$ is the distance of the squirmer from the target at time t = 0. 
\begin{figure}[t]
\centering
\includegraphics[scale = 1.2]{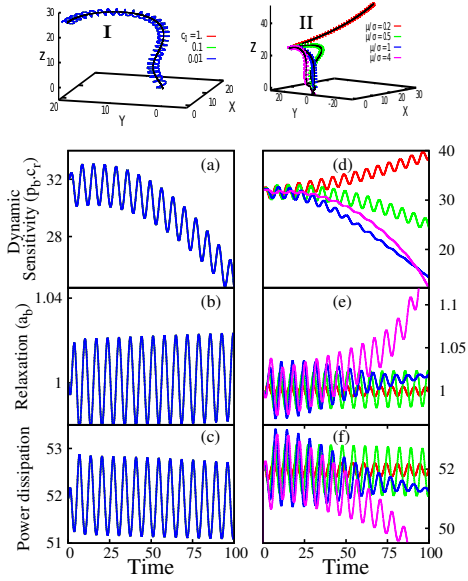}
\caption{(Color online) 
(I) Shows path of the squirmer for various strengths of chemoattractant diffusivity ($c_r$), for the radial case,
where we see that the path is independent of $c_r$.
Here, the chemical source is kept at the origin (0,0,0) and the squirmer starts at (0, 20, 25). 
The other parameters are same as in the linear case (Fig.\ref{fig:lin_nonoise}).
(a)-(c) Shows the corresponding behaviors of $p_b$, $a_b$, and power dissipation as a function of time, respectively.
(II) Shows path of the squirmer for various adaptation times for the radial case, and 
(d)-(f) are the corresponding behaviors of $p_b$, $a_b$, and power dissipation as a function of time, respectively.
Contrary to (I), here the squirmer path is influenced by the adaptation time. 
For very low value of $\mu$ (e.g., $0.2$) which is analogous to memory, the squirmer  moves away from the target.
On the other hand, very high value of $\mu$ (e.g., $4$) is also undesirable as it destroys the helical 
nature of the squirmer long before reaching the target.
Note that, sensitivity is inversely proportional to stimulus.
For higher values of $\mu/\sigma$, as the body approaches the target $p_b$ decreases and $a_b$ diverges.}
\label{fig:radtrajc}
\end{figure}

From Eq.(\ref{rad_c}) it is easy to see that the gradient $|\nabla {c}| \sim 1/r^2$ and 
the relative strength of the gradient is $|\nabla c|/c = 1/r$ which is completely independent of $c_r$. 
Whereas, in the linear case, $\nabla c$ depends on $c_1$ (see Eq.\ref{eq:l_gradient}) and $|\nabla c|/c = c_1/(c_0 + c_1 x)$, 
assuming the gradient is in the $x$ direction. 
Thus, for the radial case, the trajectory of the squirmer does not depend on the rate at which 
the chemoattractant releases from the target i.e., $c_r$, see fig. \ref{fig:radtrajc}(I). 
Sensitivity $p_b$ varies in the same manner but with different amplitudes because of the different magnitudes of $c_r$. 
We can normalize $p_b$ by multiplying it with $c_r$.
The normalization makes the amplitude of oscillation in $p_b(t)$ same for all values of $c_r$ and helps to understand that $p_b(t)$ due to different $c_r$ are in the same phase (fig. \ref{fig:radtrajc}(a)). 
The centerline of the body whirls around the target and the helical path gives rise to oscillation in $(p_b \cdot c_r)$, see fig. \ref{fig:radtrajc}(a). 
Presence of the gradient compels the body to change $a_b$ from its steady state value. 
Therefore, the body continuously winds up about the target and correspondingly $a_b$ exhibits a series of peaks, see fig. \ref{fig:radtrajc}(b). 
While the body is close to the target, $a_b$ blows up (not shown in fig.) because of 
saturation of the internal signalling network of the body due to higher chemical concentration near the target. 
As $U$ and $\Omega$ are functions of $a_b$ through the slip coefficients, they also vary over time. 
As a result, power dissipated by the body is not a constant over time but shows a sinusoidal variation, see fig. \ref{fig:radtrajc}(c). 

Like the linear case, varying $\mu$ compels the body to take different paths to reach the target, see fig.\ref{fig:radtrajc}(II). 
While for lower values of $\mu$, for example 0.2, the chemotaxis is unsuccessful as the body goes away from the target, higher values of $\mu$, 
for example 4, causes the body to loose its helicity before reaching the target, see fig.\ref{fig:radtrajc}(II). 
Therefore, the sensitivity $(p_b)$ of the body starts to increase for lower $\mu$ values as the body moves away from the target. 
For the moderate values of $\mu$, $p_b$ decreases over time but in different manners as the body takes different paths for each of the cases 
to reach the target, see fig. \ref{fig:radtrajc}(d). 
Among all the above considered $\mu$ values, only for $\mu/\sigma = 1$ the body 
reaches the target in a minimum time without loosing its helicity. 
The body looses its helicity as a consequence of saturation of the internal signaling network. 
This is manifested in the divergence of $a_b$ in fig. \ref{fig:radtrajc}(e). 
For higher $\mu$ values, $a_b$ diverges quickly implying early saturation of the network, see fig. \ref{fig:radtrajc}(e). 
The motion of the body is random after the saturation of the network and can not be explained 
with the help of Eq. \ref{eq:adt_relax}. 
For lower $\mu$ values the body is almost insensitive to the stimulus. 
As a result, the perturbation in $U$ and $\Omega$ is very low which is reflected in power dissipation curve, see fig.\ref{fig:radtrajc}(f). 

\section{Chemotaxis in presence of noise}
\label{noise}

Perfect helical motion of squirmer under a chemical gradient is highly ideal situation in general. 
There are a number of sources of noise for the squirmer in the external chemical gradient. 
For example,  
$(i)$ the releasing rate of the chemoattractant from the target can be fluctuating and may degrade with time,
$(ii)$ there can be fluctuations in the functioning of molecular motors in the axinome of the cilia,
$(iii)$ the binding of chemoattractants to the receptors of the body is a discrete random process, and
$(iv)$ the entrance of $Ca^{2+}$ in the cells is also a discrete process. 
Hence, always there is a randomness in some form or the other which influences the behavior of the system. 
While, $(i)$ is the source of external noise, $(ii), (iii), (iv)$ are the origins of internal noise. 
For simplicity, we neglect the case of $(iv)$.
Also, we assume uniform and non-degrading releasing rate of the chemoattractant from the source and smooth functioning of the molecular motors. 
This means, the cases $(i)$ and $(ii)$ are also neglected.
Consequently, the only source of randomness is the chemoattractant-receptor binding $(iii)$. 
Since the number of receptors on the body is very high, the probability that a particular chemoattractant 
will bind to a particular receptor is very low. 
Hence, the binding events can be described in terms of Poisson process.
The binding rate of chemoattractants to the receptor is $q_b(t)$. 
This is proportional to the chemical concentration at a position $\mathbf{r(t)}$ \cite{Friedrich} and is given by,
\begin{equation}
q_b(t) = \kappa\, c(\mathbf{r(t)})\,.
\label{kappa}
\end{equation}
The proportionality constant $\kappa$ can be determined from the relation \cite{berg}, 
\begin{equation}
\kappa = 4 \pi Da \, ,
\end{equation}
where $D$ is the diffusion constant of the chemoattractants and $a$ is the radius of squirmer. 
Since activation process is a stochastic one, the stimulus has a dimension of rate and not the concentration \cite{Friedrich},
\begin{equation}
\left\langle S(t) \right\rangle = q_b(t).
\end{equation}
The noise will enter in the equations through the stimulus term. 
For simplicity, we can assume that the body is in a high chemical concentration. 
In this case, the activation rate is higher than the relaxation rate i.e.  $q_b(t)> 1/\sigma$. 
The chemotactic stimulus $S(t)$ can then be replaced by the following equation \cite{Friedrich},
\begin{equation}
S(t) \approx q_b(t) + \sqrt{q_b(t)}\, \zeta (t) \,,
\label{noiseex}
\end{equation}
where $\zeta(t)$ is the Gaussian white noise with zero mean and delta correlation function, 
$\left\langle \zeta(t_1) \zeta(t_2)\right\rangle = \delta (t_1 - t_2) \,.$
The first term on RHS represents the strength of chemical gradient while the second term $\sqrt{q_b(t)}$ 
represents the strength of noise. 
For weaker chemical gradient (e.g., 1$0^{-4}$), the noise (e.g., $10^{-2}$) dominates. 
If the gradient is too shallow then the squirmer may not reach the target.

\subsubsection{Weak Concentration Gradient}

The condition for weak concentration gradient is $\nu = |\nabla c| \frac{a}{c} \ll 1$, 
where $|\nabla c|= c_1$ for the linear case and $ |\nabla c|= c_r/r^2$ for the radial case, respectively. 
Since $|\nabla c|/c$ has a dimension of inverse length, $\nu$ is a dimensionless parameter. 

\subsubsection{Strong concentration gradient}

The condition for strong concentration gradient is $\nu \gg 1$. 
Higher the value of $c_1$ and $c_r$ stronger the gradient is. 
However, values of $c_1 >1$ and $c_r > 1$ are not allowed because the equations used to describe adaptation and relaxation 
dynamics (Eq.~\ref{eq:adt_relax}) are valid in the limit of weak concentration gradient only.
\begin{figure}[ht]
\centering
\includegraphics[scale = 0.6]{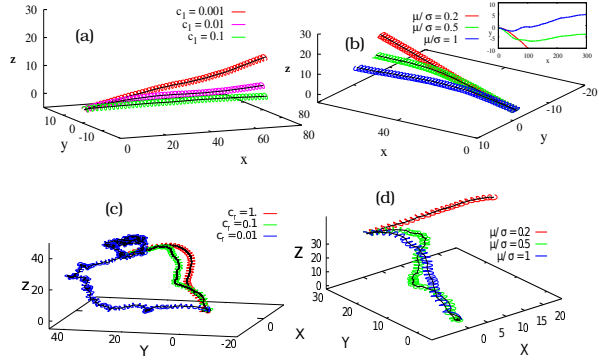}
\caption{(Color online)
Path of the squirmer in presence of noise.
(a) Shows the path for linear case. We have set $\mu/\sigma = 1.2$. 
The rest of the parameters are as in fig. (\ref{fig:lin_nonoise}).
The strength of noise varies as $\sqrt{\kappa c}$,  where we have chosen $\kappa \approx 10$. 
The squirmer moves in a noisy helical path. 
The fluctuations are reduced for higher values of concentration as shown in the figure. 
(b) Shows the path, for the linear case, for various values of $\mu/\sigma$, where we have set the concentration gradient $c_1 = 0.01$. 
Here, varying adaptation time does not alter the effect of noise. The body moves in an irregular manner in the $xy$ plane. 
(c) Shows path for the radial case, for various strengths of chemoattractant diffusivity ($c_r$).
Interestingly, while for noise free case the movement of squirmer was independent of $c_r$, 
for the noise case it depends on $c_r$. 
(d) Shows path for the radial case, for various values of $\mu/\sigma$, where we have set $c_r = 0.1$.
Here, noise does not offer much advantage except making the helical path noisy.}
\label{nctrajc}
\end{figure}

Noise plays a dominant role in the weak chemical concentration gradient limit. 
In the linear case, with increasing gradient, the trajectory of the squirmer becomes less noisy and 
aligns quickly in the direction of the gradient, see fig. \ref{nctrajc}(a). 
On the other hand, weak memory i.e., 
low $\mu$ causes unsuccessful chemotaxis and noise does not make this situation any better, 
path becomes irregular following the gradient, see fig.\ref{nctrajc}(b).  
In general, activation rate is higher for stronger gradient and is favorable for $Ca^{2+}$ 
entrance in the cells which lead to an effective change in the linear velocity and the rotation rate of the body. 
If the gradient is very weak the squirmer will never be able to make its axis parallel to the direction of the gradient. 
Similarly, if the adaptation time is low then squirmer cannot effectively sense the chemical 
gradient and as a result it may not follow the gradient, see the inset of fig. \ref{nctrajc}(b).

For the radial case, noise makes the squirmer to depend on $c_r$. 
The relative strength of stimulus over the body is, $|\nabla S|/S = -(c_r/r^2 + \sqrt{c_r/r^3}\zeta(t)/2)/(c_r/r + \sqrt{c_r/r} \zeta(t))$ 
which controls the trajectory of the body. This is not independent of $c_r$. 
Note, that for higher $c_r$ noise is less. 
This is clearly visible in fig. \ref{nctrajc}(c), where the trajectory is smoother for higher $c_r$.
If $c_r$ is sufficiently low then it is possible that the noise may becomes excessively dominating and the body may not reach the target. 
Analogous to the linear case, low adaptation time also diminish the success rate of the squirmer (see fig.\ref{nctrajc}(d)). 
Whenever the time scale of noise ($t_{Noise}$) is comparable to the adaptation time, 
the system is dominated by the former for both the linear and radial cases, as discussed in subsequent section.
The body remains insensitive to the gradient and drive away from the target for lower $\mu$, 
as shown in figs \ref{nctrajc}(b) and (d). 
For higher $\mu$, the body gets more time to sense the local stimulus level. 

Furthermore, for the linear case the considered $\mu$ values are much larger compared to $t_{Noise}$ and for 
the radial case the situation is reverse. As a result, the noise is subtly reduced in $p_b$ with increasing $\mu$, 
see fig.\ref{lin_noise_sup} and \ref{rad_noise_sup}.
Because of this reason, changing adaptation time can not reduce the irregularity in different properties of 
the squirmer like power dissipation and relaxation visibly (see graphs in appendix). 

\subsection{Mean first passage time}

\begin{figure}[h]
\centering
\includegraphics[scale = 1.4]{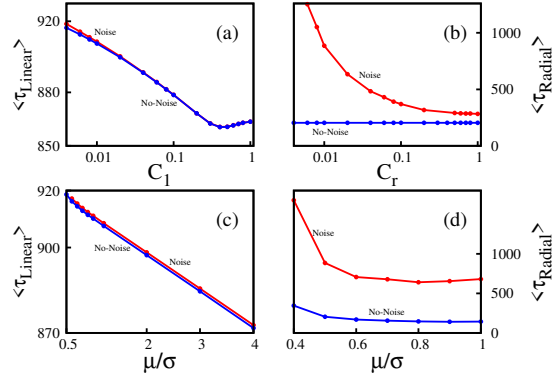}
\caption{(Color online) Mean first passage time for the linear (a) and the radial (b) cases have been obtained numerically. 
For the linear case, $\uptau_{Linear}$ is the time at which the body crosses $l \geq 200a$ on the $x-$axis, where $a$ is the radius of the body. $\mu/\sigma$ has been fixed to $1.2$ here. The other parameters have the same values as in the fig. \ref{fig:lin_nonoise}(I).  
For the radial case, $\uptau_{Radial}$ is the time taken by the body to travel across $r_{0} = a$, 
where $r_{0}$ is the distance from the source which is located at the origin $(0,0,0)$. $\mu/\sigma$ has the value $0.5$ here. For the other parameters we have considered the same values as in the fig.\ref{fig:radtrajc}(I). 
The mean first passage time $\uptau_{Linear/Radial}$ decreases with increasing strength of 
the concentration gradient or chemoattractant diffusivity. 
This implies that the effect of noise on the motion of the body is reduced for  higher concentration gradient or chemoattractant diffusivity. 
(c) $\uptau_{Linear}$ decreases with increasing $\mu$ for both noise free and noise case. $c_1$ has been fixed to $0.01$ in this case and other parameters have the same values as in the fig.\ref{fig:lin_nonoise}(II).  
(d) The nature of variation of $\uptau_{Radial}$ is same for both in the presence and absence of noise. 
Only difference is that the values are higher in presence of noise. We have set $c_r = 0.1$ here. The other parameters have the same values as in the fig.\ref{fig:radtrajc}(II).}
\label{diffusivity}
\end{figure}

To understand the strength of coupling of noise to chemotaxis, we have numerically calculated the 
first passage times ($\uptau$), see fig. \ref{diffusivity}.
It is the time taken by the body to reach the target, radial case, or to cross a reference point in space, linear case, for the first time. 
\begin{subequations}
\begin{equation} 
\uptau_\mathrm{Linear} = inf\{t \geq 0; x(t) \geq l\} \,,
\end{equation}
\begin{equation}
\uptau_\mathrm{Radial} = inf\{t \geq 0; r(t) \leq r_0\} \,,
\end{equation}
\end{subequations}
where $\uptau_\mathrm{Linear}$ and $\uptau_\mathrm{Radial}$ are the first passage times for the linear and the radial cases, respectively. 
Here, $l$ is a point on the $x-$ axis ($l >> x(0)$) and $r_0$ is a point very near to the chemical source ($r_0 \le a$). 
Since noise makes the process stochastic, we have calculated ensemble averaged first passage times.

In absence of noise, depending on the strength of the gradient the body takes different times to reach the reference point. 
Weaker gradient causes the body to take longer time compared to that corresponding to stronger gradient, see fig.\ref{diffusivity}(a). 
On the other hand, the relative strength of the gradient is independent of $c_r$ for the radial noise 
free case which makes the body to take same route to reach the target. 
Hence, it takes same time to reach the source, see fig.\ref{diffusivity}(b). 
However, for a fixed gradient the choice of $\mu/\sigma$ values decides the success of chemotaxis. 
As it was discussed before, for low values of $\mu$, the body takes a longer path to reach the target. 
The same reflected in $\uptau_\mathrm{Linear/Radial}$, see fig. \ref{diffusivity}(c) and (d). 
On the contrary, higher $\mu$ values cause gain in the linear velocity of the body, 
as a result the body takes lesser time to reach the reference point, see fig.\ref{diffusivity}(c).
In the radial case we did not consider the mean first passage time for $\mu/\sigma$ beyond $1$ because in that 
region the squirmer looses its helicity even in the absence of noise resulting from the saturation of internal signalling network. 

In presence of noise the nature of the $\uptau_\mathrm{Linear/Radial}$ highly depends on the characteristic timescale associated with the noise, i.e. $t_{noise} = 1/q $. 
For the linear case, $q = \kappa (c_0 + c_1 x)$, and for the radial, $q = \kappa (c_r/r)$. 
Here for the given initial conditions, maximum available value for $t_{noise}$ is, $0.1$ for the linear case and $\approx 32$ for the radial case. 
As the body starts to approach the source $t_{noise}$ decreases. 
Note that for the linear case, $t_{noise}$ is much smaller compared to the considered timescales associated with the relaxation and adaptation dynamics, i.e. $(\mu/\sigma)$. 
Hence, the effect of noise is not pronounced, see figs. \ref{diffusivity} (a) and (c).
However, for a very low $c_1$ noise may show some impact, see fig. \ref{diffusivity} (a).   
Whereas in the radial case, $t_{noise}$ is much greater compared to the considered values of 
$\mu/\sigma$ resulting in high $\uptau_{Radial}$ over the noise free situation, see figs. \ref{diffusivity} (b) and (d).

\section{Conclusions}

In the current work, we have considered the chiral squirmer model to study the chemotaxis with both the linear and the radial chemical gradients. 
An important feature of this model is the rotational degree of freedom which is an advantage in the process of chemotaxis. 
For example, if the direction of the chemical gradient is perpendicular to the motion of the achiral squirmer (without rotational motion), 
the body will not be able to follow up the gradient as there is no component of rotational motion which will help it to 
rotate its polar axis towards the direction of the gradient. 

We have used Eq.(\ref{eq:adt_relax}) to describe the adaptation and relaxation mechanism of the body 
in presence of an external chemical gradient.
The body changes its course of motion and this totally depends on the relative strength of the chemical gradient. 
Whereas, for the linear case it is a function of $c_0$ and $c_1$ (strength of the gradient), 
for the radial case it is independent of $c_r$ (chemoattractant diffusivity). 
As a result, for the linear case, higher values of $c_1$ leads to sharper bending of the path while 
varying $c_r$ has no effect on the trajectory of the body for the radial case.

Note that chirality is not the sole parameter behind the successful chemotaxis.  
The adaptation time $\mu$ also plays a vital role in the process of chemotaxis and 
proper choice of $\mu$ optimize the time, at which the body reaches the target (radial case) or align its direction 
of motion in the gradient direction (linear case). 
Because of this, in reality, a subpopulation of a colony of microorganisms reach the target quickly than others. 
$\mu$ is analogous to the memory for the given system. That is why, it really does not matter 
how strong the gradient is, lower $\mu$ value i.e., weak memory will always give rise to unsuccessful chemotaxis. 

In reality, noise effects the process of chemotaxis. 
In the presence of noise, body receives a stimulus which fluctuates over time and over the surface of the body. 
The response of the body to this stimulus also becomes noisy which is reflected in its irregular path. 
In short, noise plays a dominant role in the weak concentration regime. 
As a result the body cannot align its path in the direction of the gradient, for the linear case, and 
may not reach the target, for the radial case. 
For a comparatively strong concentration, the effect of noise is suppressed to a great extent. 
Hence, the system is more ordered which is reflected in the behavior of mean first passage time $\uptau_{Linear/Radial}$. 
It diverges when $c_1$ or $c_r$ is very small, implying arbitrary movement of the body in spite of 
the presence of the concentration gradient. 
Hence, $c_1$ or $c_r$ needs to be high to drive the body towards the target effectively.
This study is very useful to understand the chemotactic behavior of ciliated microorganisms and also to design synthetic bodies 
for targeted applications, e.g., drug delivery, wound healing, etc.

\appendix

\section{}

\begin{figure}[htb]
\includegraphics[scale = 1.3]{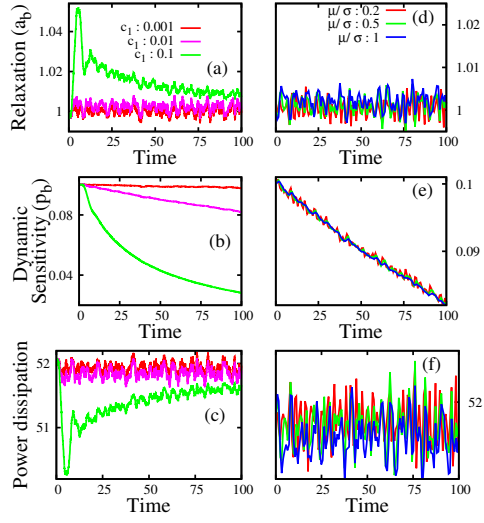}
\caption{(Color online) (a)-(c) The variation of $p_b$, $a_b$ and power dissipation over time 
under a linear gradient are the same as those for noise-free case. 
The only difference here is that, those variations are not smooth but rather noisy. 
Increasing strength of the gradient makes the peaks in $a_b$ more prominent and decay in $p_b$ steeper. 
For a weaker gradient, $a_b$ keeps fluctuating about its steady state value so is the motion of the body. 
(d)- (f) For varying $\mu$, the response of the body is more or less the same. The peaks are irregular in $a_b$ due to noise. 
The same goes for the power dissipation curve also. Only the $p_b$ is less noisy for higher $\mu$. }
\label{lin_noise_sup}
\end{figure}

\begin{figure}[htb]
\centering
\includegraphics[scale = 1.3]{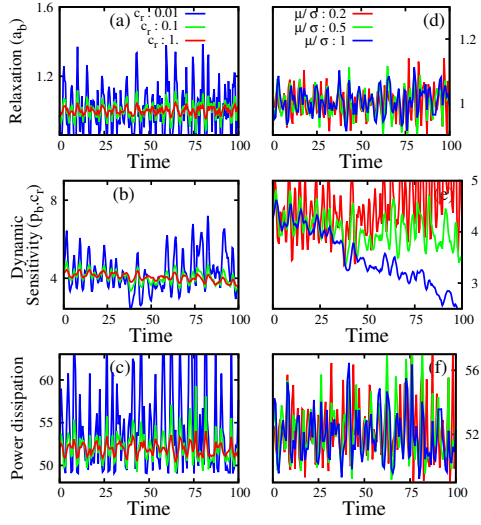}
\caption{(Color online)(a)-(c) The variation in $a_b$, $p_b$ and power dissipation under a radial gradient is 
highly dependent on the value of $c_r$. Increasing $c_r$ suppresses the effect of noise. 
Here also like in the linear case, the peaks in $a_b$, $p_b$ and power dissipations are irregular. 
(d)-(f) For varying $\mu/\sigma$, the gross nature of $a_b$, $p_b$ and power dissipation are not 
changed much because of noise as compared to that for the noise-free case.  }
\label{rad_noise_sup}
\end{figure}

In section \ref{noise}, we have investigated the behavior of squirmer in the presence of a chemical gradient under the influence of noise. 
The parameters related to the adaptation and relaxation mechanism show a noisy behavior. 
Consequently, the induced random motion gives rise to fluctuating behavior in other parameters, e.g. adaptation, relaxation, and power dissipation. 
The noisy behavior of all these parameters are depicted for varying gradient, $c_1$ (linear case) see fig.\ref{lin_noise_sup} (a)-(c), 
and varying chemoattractant diffusivity, $c_r$ (radial case), see fig. \ref{rad_noise_sup} (a)-(c). 
We have observed that varying $\mu/\sigma$ does not alter the effect of noise for a fixed $c_1$, see fig.\ref{lin_noise_sup}(d)-(f), 
and $c_r$, see fig. \ref{rad_noise_sup}(d)-(f).

\end{document}